\documentclass[convention,peer-reviewed]{aesconf} 

\graphicspath{{./}{figs/}}

\usepackage[utf8]{inputenc}

\usepackage{microtype}
\makeatletter
\providecommand\MT@suspend@tagging{}
\providecommand\MT@resume@tagging{}
\makeatother

\usepackage[numbers,square]{natbib}

\usepackage{booktabs}
\usepackage{color}
\usepackage{url}

\usepackage{dsfont} 
\usepackage{amsmath,amssymb,amsfonts,bm}

\usepackage{algorithm}
\usepackage[noend]{algpseudocode}

\usepackage{subfig} 

\newcommand{\field}[1]{\mathbb{#1}} 

\newcommand{\BIGO}[1]{\textrm{O} \left ( {#1} \right )}

\newcommand{\BRAK}[1]{\left [ {#1} \right ]}
\newcommand{\CBRAK}[1]{\left \{ {#1} \right \} }
\newcommand{\ABS}[1]{\left | {#1} \right | }
\newcommand{\PAREN}[1]{\left ( {#1} \right )}

\newcommand{\argmax}[1]{\underset{#1}{\operatorname{arg}\,\operatorname{max}}\;}
\newcommand{\argmin}[1]{\underset{#1}{\operatorname{arg}\,\operatorname{min}}\;}

\newcommand{\NORM}[1]{\left \| {#1} \right \| }

\newcommand{\FLOOR}[1]{\left \lfloor {#1} \right \rfloor}

\newcommand{\VEC}[1]{\boldsymbol{#1}}
\newcommand{\MAT}[1]{\boldsymbol{#1}}

\newcommand{\RANGEW}[2]{\BRAK{{#1} \textrm{ to } {#2} }}

\newcommand{\plus}{\raisebox{0\height}{\scalebox{.55}{+}}}

\DeclareMathSymbol{\minus}{\mathbin}{AMSa}{"39}

\algnewcommand\algorithmicinput{\textbf{Input:}}
\algnewcommand\INPUT{\item[\algorithmicinput]}
\algnewcommand\algorithmicoutput{\textbf{Output:}}
\algnewcommand\OUTPUT{\item[\algorithmicoutput]}

\title{Fast Time-Varying Exponentiated Convolution Methods for Generative Direction Dependent Reverberation}

\author[1]{Yuancheng Luo}

\affil[1]{NuSpace Audio, Cambridge, MA, USA}

\correspondence{Yuancheng Luo}{luoyuancheng@gmail.com}

\lastnames{Luo}

\shorttitle{Exponentiating Direction Dependent Reverberation}

\begin{document}

\twocolumn[
\maketitle 

\begin{onecolabstract}

Spherical harmonic encoded acoustic sound-fields capture directional characteristics of room impulse responses that are useful for accurate spatial audio reproduction. However, high costs of multi-microphone measurements and numerical simulations motivate alternative data-set augmentation and synthetic data generation methods that supplement small collections. This paper introduces time-varying exponentiated convolution methods that transform both Gaussian noise and impulse responses into reverberation and modified spectral-decay fields respectively. We derive two recursive and fast convolution algorithms that extend into the spherical harmonic domain, model smooth reverberation time distributions with non-stationary Gaussian processes, and realize an optimal filter design. Experiments evaluate computational performance, and validate out-of-distribution generated impulse responses.

\end{onecolabstract}
]

\section{Introduction}
\label{sec:intro}

Spatial room impulse responses (SRIRs) that decompose acoustic fields over bounded spaces have found numerous applications in spatial audio reproduction over both physical and virtual loudspeaker arrays for auralization. In optimal experimental design, \textit{hardware auralization} of microphone and loudspeaker topology enables  performance evaluations on various downstream processors such as beam-forming, acoustic tuning, speech enhancement, and sound-source localization. Spherical harmonic (SH) \cite{muller2006spherical, rafaely2015fundamentals, jarrett2017theory} formatted SRIRs such as Ambisonics \cite{zotter2019ambisonics}, are well-suited for auralization as their rotational properties and separability from the hardware's geometry increase the modeling capacity of diverse experimental conditions; SRIRs can be integrated with freely rotated plane-wave decompositions of a transducer's free-field acoustic responses. However, acquiring large and varied SRIR data-sets remains a challenge due to practical costs of measurement and numerical simulation. 
We therefore investigate synthetic and data augmentation methods of SH-SRIRs as to complement data collection efforts:

Approximate RIR models such as SH domain image-source method \cite{wang2023time, xu2024simulating, luo2021FSRR} are efficient alternatives to expensive numerical simulations, capable of generating large quantity of SH-RIRs by varying simplified shoe-box room geometries and wall-material absorption properties. Time-varying smoothing kernels \cite{hamilton2021air, fagerstrom2022dark} and low-pass filters \cite{kates2020adding, valimaki2017late} can augment simulated RIRs with spectral-decay characteristics of distance-based atmosphere attenuation. Directional filtering using SH filter-banks \cite{zotter2012all, hold2021spatial} can modify the directivity of sound-fields. 
Our work synthesizes these approaches by generalizing time-varying filtering for exponentiating finite-impulse-response (FIR) kernels that control for direction dependent spectral-decay under SH expansions. As a result, we can both augment and generate SH-SRIRs with direction dependent spectral-decay distributions sampled from a Gaussian process (GP) \cite{rasmussen2003gaussian}.

Our paper is organized as follows: Section \ref{SEC:TIME_VAR_CONV} introduces two time-varying exponentiating convolution methods for modifying IR spectral-decay without adding delay. Section \ref{SEC:DIR} extends our methods into the SH domain for directional filtering and reverberation; we present a filter optimization program and a non-stationary GP model of reverberation time distributions. Section \ref{SEC:EXP} evaluates several experiments with reverberation time sampling, filter fitting sensitivity, and SH-SRIR generation. Section \ref{SEC:CONC} concludes the work. Our implementation is available online\footnote{\url{https://github.com/yluo1/SH-IRT}}.

\section{Exponentiating Convolutions}
\label{SEC:TIME_VAR_CONV}

Let us consider the following generalization of the convolution function $f(\VEC{h}, \, \VEC{g})$  between a fixed FIR $\VEC{h}$ and a time-varying FIR filter $\VEC{g}(m)$ undergoing exponential-convolution at each time index $m$:
\begin{equation} \label{EQ:FILTER_DEF}
\displaystyle
\begin{split}
\VEC{h}  =  \BRAK{h_1, \hdots, h_M} \in \field{R}^{M}, \quad
\VEC{g} = \BRAK{g_1, \hdots, g_N} \in \field{R}^{N}, \\
\VEC{g}(m)  = \VEC{g}(m - 1) \circledast \VEC{g} =  \circledast_{k=1}^m \, \VEC{g} \in \field{R}^{m(N \minus 1) \plus 1}, 
\end{split}
\end{equation}
where $\circledast$ is the convolution operator.
The $n^{th}$ sample of the time-varying convolution function $f(\VEC{h}, \, \VEC{g})$ follows
\begin{equation} \label{EQ:EXP_CONV}
\displaystyle
\begin{split}
f[n] & = \sum_{m = 1}^{MN} h[n - m + 1] \, g(n -  m + 1) [m], \\
\end{split}
\end{equation}
where its non-zero array elements are indexed via
\begin{equation} \label{EQ:AR_NOTES}
\displaystyle
\begin{split}
h[m]  & = \left \{  \begin{array}{cc}
 h_m, & 1 \leq m \leq M \\[3pt]
 0, & \textrm{Otherwise}
 \end{array} \right . , \quad 
\textrm{Fixed FIR}
 \\ 
 g[n]  & = \left \{  \begin{array}{cc}
 g_n, & 1 \leq n \leq N \\[3pt]
 0, & \textrm{Otherwise}
 \end{array} \right . , \, \, \, \, \, \quad 
 \textrm{Exp. Filter} \\
 g(m)[n]  & = \left \{ \begin{array}{cc} 
\sum_{k = 1}^{ m N } g[n - k + 1] \, g(m  - 1) [k],  & m > 1 \\[3pt]
g[n], & m = 1 \\[3pt]
0, & m < 1
\end{array} \right . .
\end{split}
\raisetag{18ex}
\end{equation}
The later samples in $\VEC{h}$ are convolved with samples of higher-order filters $\VEC{g}(m)$ under exponentiation. Thus, direct computation of $f(\VEC{h}, \, \VEC{g})$ can be expressed in terms of summations over sample-weighted exponentiating convolutions in Algorithm \ref{ALG:EXP_CONV:DIRECT}. Its asymptotic run-time costs $\BIGO{M^2 N}$ however are prohibitive  as the exponentiating filter $\VEC{g}(m)$ grows by $N-1$ samples per time-step, yielding a size of $M(N-1)+1$ when directly convolving with filter $\VEC{h}$ in the time-domain.

\begin{algorithm}
\caption{Direct Exponentiating Convolution}\label{ALG:EXP_CONV:DIRECT}
\begin{algorithmic}[1]
\Procedure{DirExpConv}{$\VEC{h}$,  $\VEC{g}$} 
   \INPUT $\VEC{h} \in \field{R}^{M}$ \quad \qquad Fixed FIR
   \INPUT $\VEC{g} \in \field{R}^{N} $ \, \, \,  \qquad Exponentiating FIR
   \OUTPUT $\VEC{f} \in \field{R}^{MN}$ \quad \ \ \ Accumulating FIR
   \State $\VEC{f} \gets \VEC{0}^{MN} $ 	\Comment{Initialize accumulator}
   \State $\VEC{u} \gets  \VEC{g}$ 	\Comment{Initialize Exp. FIR}
   \For{$m = 1 \dots M$ } \Comment{Sum convolutions over $\VEC{h}$}
       \State $f\RANGEW{m}{m N}  \gets f \RANGEW{m}{m N} +   h[m] \,  \VEC{u} $
       \State $\VEC{u} \gets \VEC{u} \circledast \VEC{g}$ \Comment{Exp. convolution via Eq. \eqref{EQ:FILTER_DEF}}
	\EndFor
	\Return $\VEC{f}$
\EndProcedure
\end{algorithmic}
\end{algorithm}

Fortunately, we can obtain a faster algorithm by deriving the transfer functions of both $\VEC{g}(m)$ and  $f(\VEC{h}, \VEC{g} )$. The transfer function $G(m, z) = G^m(z)$ of the exponentiating filter $\VEC{g}(m)$ over the $Z$ domain is the exponentiated transfer function $G(z)$ of $\VEC{g}$. The transfer function $F_{M}(z)$ of $f(\VEC{h}, \VEC{g} )$ following Algorithm \ref{ALG:EXP_CONV:DIRECT} can be expressed by
\begin{equation} \label{EQ:EXP_CONV_TF}
\displaystyle
\begin{split}
F_{M}(z)  & = \sum_{m=0}^{M \minus 1}  h_{m \plus 1} z^{\minus m} G^{m \plus 1}(z)  , \quad \textrm{Time-vary Conv.} \\
G^{m}(z) &  = \PAREN{\sum_{n=0}^{N \minus 1} g_{n \plus 1} z^{\minus n}}^{m}, \qquad \quad \,\, \textrm{Exp. Filter}
\end{split}
\raisetag{5ex}
\end{equation}
where we observe that the exponentiated filter's transfer function  $G^{m}(z) = G^{\FLOOR{m/2}}(z) G^{m \minus \FLOOR{m/2}}(z)$ can be recursively divided into halves of $m$ via the floor function $\FLOOR{*}$. Moreover, the transfer function $F_M(z)$ can be recursively divided into upper and lower sample halves of $\VEC{h}$, and expressed as the following recurrence relation:
\begin{equation} \label{EQ:EXP_CONV_TF_RECUR}
\displaystyle
\begin{split}
F_{M}(z, \, \VEC{h}) & = \sum_{m=1}^{M} h[m]  z^{\minus (m  \minus 1)} G^{m}(z) \\
&  = \left \{  \begin{array}{cc}
 \underline{F_{ \underline{M} }}(z) + \overline{F_{ \underline{M} }}(z)  \, G^{\underline{M}}(z)  \, z^{\minus \underline{M} }  , &
M > 1 \\[4pt]
h[m] G(z), & 
M = 1 
\end{array} \right . , \\
\overline{F_{ \underline{M} }}(z) & =   F_{M \minus \underline{M}} \PAREN{z, \, h \BRAK{ \PAREN{\underline{M} + 1}, \, \hdots, \, M } }, \\[2.5pt]
\underline{F_{ \underline{M} }}(z) & = F_{ \underline{M} } \PAREN{z, \, h\BRAK{1, \,  \hdots, \, \underline{M}  }}, \quad
\underline{M} = \FLOOR{\frac{M}{2}},
\end{split}
\raisetag{4ex}
\end{equation}
where $F_{M}(z, \, \VEC{h})$ decomposes into summations of two transfer functions over $\VEC{h}$ separated w.r.t. pivot index $\underline{M}$. The lower-half transfer function $\underline{F_{ \underline{M} }}(z) $ remains unchanged, whereas the upper-half $\overline{F_{ \underline{M} }}(z)$ is multiplied by the exponentiated filter's transfer function $G^{\underline{M}}(z)$ and delay $ z^{\minus \underline{M}}$. The latter constitutes the anticipated work required to recursively compute the transfer function when sub-divided upto a maximum depth of $\log_2(M)$.

\begin{algorithm}
\caption{Optimized Exponentiating Convolution }\label{ALG:EXP_CONV:OPT}
\begin{algorithmic}[1]
\Procedure{OptExpConv}{$\VEC{h}$,  $\VEC{g}$} 
   \INPUT $\VEC{h} \in \field{R}^{M}$ \quad \qquad Fixed FIR
   \INPUT $\VEC{g} \in \field{R}^{N} $ \, \, \,  \qquad Exponentiating FIR
   \OUTPUT $\VEC{f} \in \field{R}^{MN}$ \quad \ \ \ Accumulating FIR
   \State $\underline{M} \gets \FLOOR{\frac{M}{2}}$
   \vspace{0.25em}
   \If{$M \leq M_{min}$} \Comment{$M_{min}$ is nominally small}
   		\State $\VEC{f} \gets $ DirExpConv($\VEC{h}$, $\VEC{g}$) \Comment{via Algo. \ref{ALG:EXP_CONV:DIRECT}}
   \Else \Comment{Compute two halves via Eq. \eqref{EQ:EXP_CONV_TF_RECUR} }
		\State $\overline{\VEC{f}} \gets $ OptExpConv($h\RANGEW{\underline{M} + 1}{M}$, $\VEC{g}$)
   		\State $\underline{\VEC{f}} \gets $ OptExpConv($h\RANGEW{1}{\underline{M}}$, $\VEC{g}$)   
   		\State $\VEC{f} \gets \BRAK{\underline{\VEC{f}}, \, \VEC{0}^{(M \minus \underline{M}) N } } + \BRAK{\VEC{0}^{\underline{M}}, \,  \overline{\VEC{f}}  \circledast \,  \VEC{g}(\underline{M})  } $ 
   \EndIf
	\Return $\VEC{f}$
\EndProcedure
\end{algorithmic}
\end{algorithm}

\begin{figure*}[h]
\centering 
    \subfloat[Input Gaussian noise $\VEC{h}$ to output $f(\VEC{h}, \VEC{g})$ spectrograms \label{FIG:EXP_CONV:SAMP}]{%
        \includegraphics[width=0.49\textwidth]{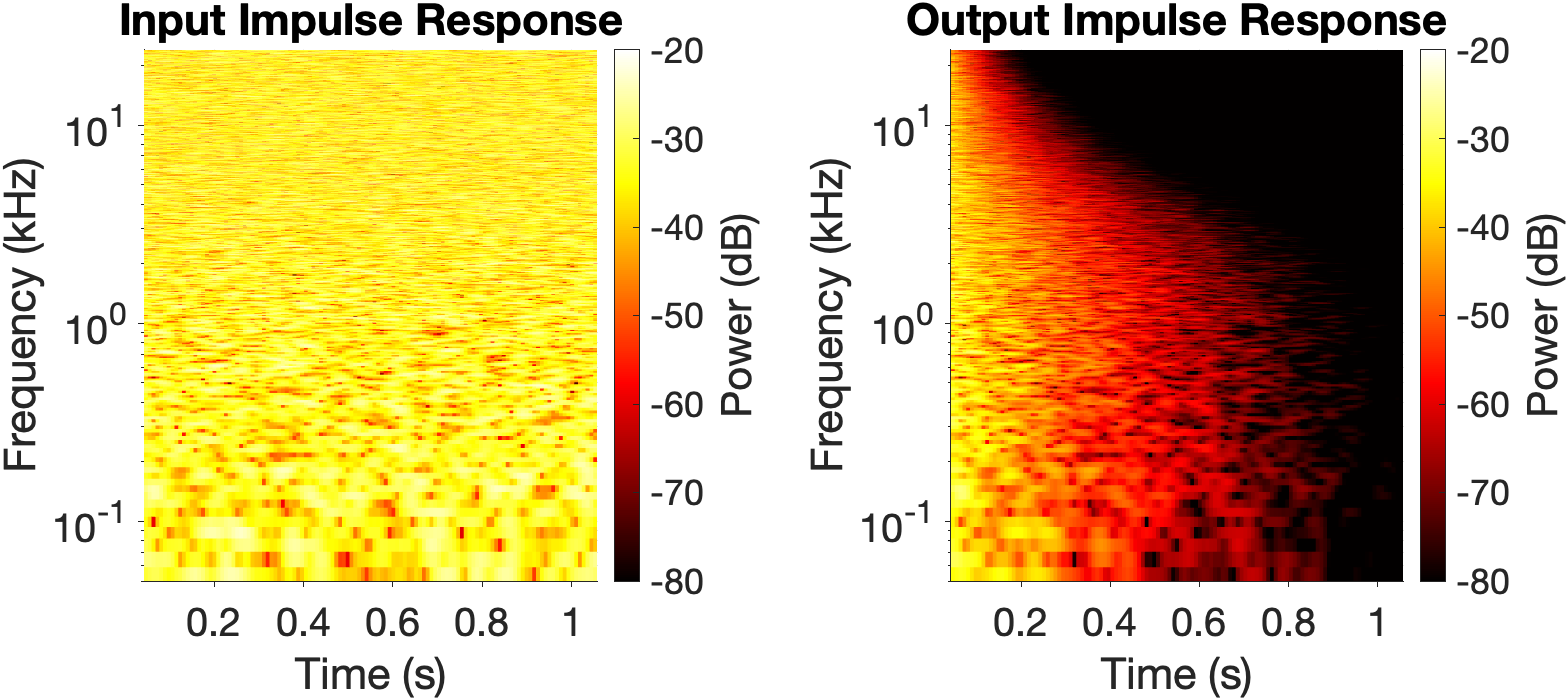}%
    }
    \hspace{1pt}
    \subfloat[Transform spectral-decay of input RIR $\VEC{h}$ to output $f(\VEC{h}, \VEC{g})$\label{FIG:EXP_CONV:SAMP_WAV}]{%
        \includegraphics[width=0.49\textwidth]{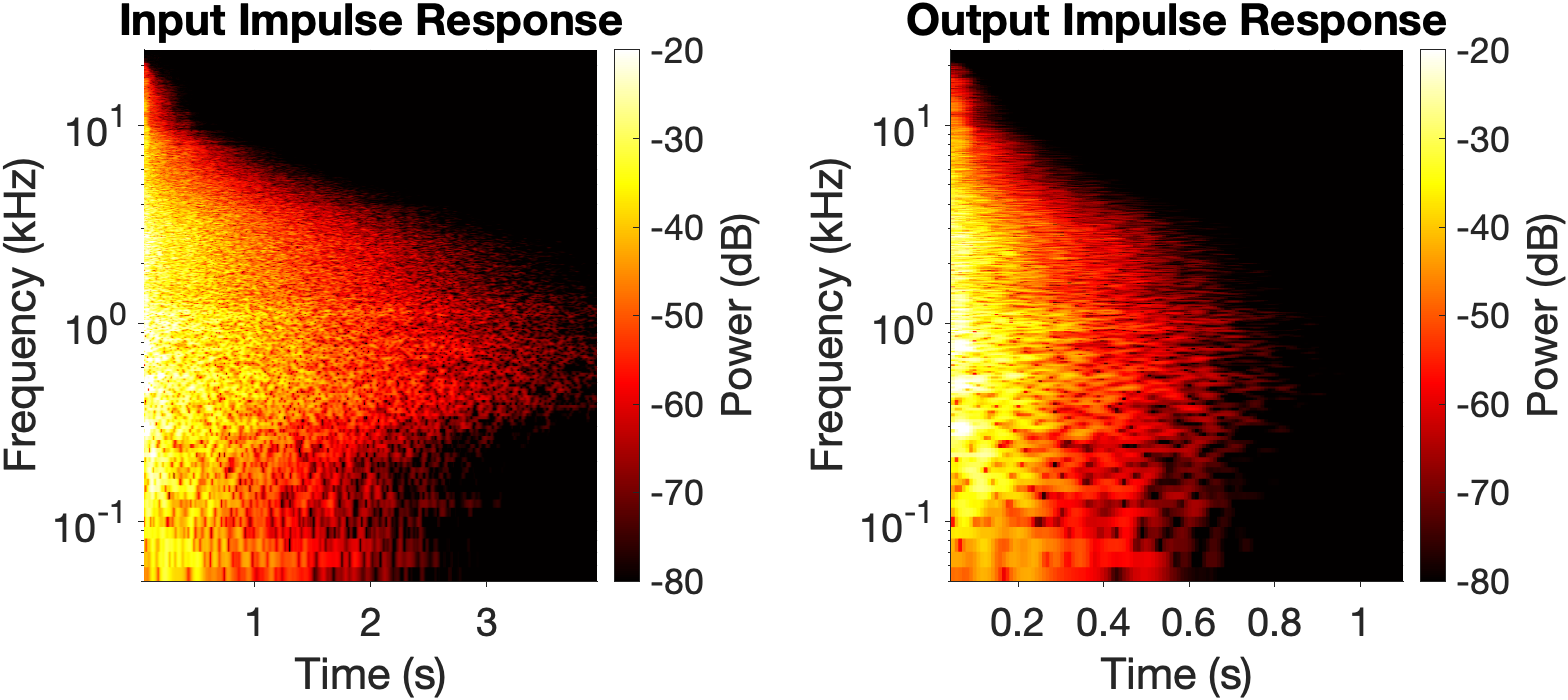}%
    }
      \caption{\label{FIG:EXP_CONV}Exponentiating filtering convolves the fixed input FIR $\VEC{h}$ with the time-varying filter $\VEC{g}$ undergoing sample-wise exponential-convolution of Eqs. \eqref{EQ:FILTER_DEF}, \eqref{EQ:EXP_CONV} to achieve target frequency dependent attenuation rates.}
\end{figure*}

We implement the recurrence relation of Eq. \eqref{EQ:EXP_CONV_TF_RECUR} in our optimized exponentiating convolution Algorithm \ref{ALG:EXP_CONV:OPT}. The computational work in each  recursive call is the convolution operation between $\overline{\VEC{f}}$ and $\VEC{g}(\underline{M})$, which can be efficiently performed in the frequency domain via the Fast Fourier transform (FFT) at $\BIGO{MN \log (MN) }$ computation costs; nominally small partition sizes revert to the direct method in Algorithm \ref{ALG:EXP_CONV:DIRECT} when its costs are cheaper than the FFT implementation. 
Furthermore, the problem sub-divides into equal-sized partitions which simplifies the complexity analysis; integrating the work over all partitions and depths gives the total asymptotic computation cost of $\BIGO{MN \PAREN{ \PAREN{ \log (M) }^2 + \log(M) \log(N) }}$. Note that in practice,  $N \lll M$ as we design filters $\VEC{g}$ to be near minimum-phase with magnitude responses that produce expected reverberation time profiles. The examples in  Fig. \ref{FIG:EXP_CONV} use a $N = 4$ order filter $\VEC{g}$ to shape frequency dependent attenuation rates of an $M = 72000$ length Gaussian noise $\VEC{h}$ in Fig. \ref{FIG:EXP_CONV:SAMP}, and $M=192000$ length RIR in Fig. \ref{FIG:EXP_CONV:SAMP_WAV}.
Next, we present an alternative but more expensive auto-regressive formulation of the exponentiating convolution process.

\textbf{Recursive Auto-regressive Convolution:} 
Consider the modified exponentiating convolutions of Eq. \eqref{EQ:EXP_CONV}, by which $f[n]$ are samples of a first-order auto-regressive function $f_m[n]$ where  $f[n]  = f_{M}[n]$. The auto-regressive function's recurrence relations are given by
\begin{equation} \label{EQ:REC_CONV}
\displaystyle
\begin{split}
f_{m}[n]   = \left \{ \begin{array}{cc}
f_{m \minus 1}[n], & m > 1 , \,  n < m \\[6pt]
\sum_{k = m}^{n} f_{m \minus 1}[k] \, g[n - k + 1],  & m > 1, \, n \geq m \\[6pt]
\sum_{k = 1}^{M} h[k] \,  g[n - k + 1], & m = 1 \\[6pt]
h[n], & m = 0
\end{array} \right . ,
\end{split}
\raisetag{5ex}
\end{equation}
where the initial state of $f_0[n]$ is FIR $\VEC{h}$, and current samples of $f_m[n]$ are convolutions of partial samples $f_{m \minus 1}[n]$ from the previous time-step with filter $\VEC{g}$. We give the direct implementation in Algorithm \ref{ALG:RECUR_CONV:DIRECT} where  a sliding window of $f[n]$ recursively convolves with $\VEC{g}$.

\begin{algorithm}
\caption{Direct Recursive Convolution}\label{ALG:RECUR_CONV:DIRECT}
\begin{algorithmic}[1]
\Procedure{DirRecurConv}{$m$, $\VEC{h}$,  $\VEC{g}$} 
   \INPUT $m$ \qquad \qquad \qquad \quad \   Trim index
   \INPUT $\VEC{h} \in \field{R}^{M}$ \quad \qquad \qquad Fixed FIR
   \INPUT $\VEC{g} \in \field{R}^{N} $ \, \, \,  \qquad  \qquad Exponentiating FIR
   \OUTPUT $\VEC{f} \in \field{R}^{M \plus m(N \minus 1) }$  \quad  \ \ Accumulating FIR
   \State $\VEC{f} \gets \VEC{0}^{M \plus m(N \minus 1) } $  \Comment{Initialize accumulator}
   \State $f \RANGEW{1}{M} \gets  \VEC{h}$ \Comment{Initialize head of $\VEC{f}$ to $\VEC{h}$}
   \For{$i = 1 \dots m$ } \Comment{Conv. $\VEC{g}$  over trimmed $\VEC{f}$}
   	   \State $\VEC{u} \gets f \RANGEW{i}{M +(N - 1)(i - 1)} \circledast  \VEC{g}$
       \State $f\RANGEW{i}{M + (N - 1)i}  \gets \VEC{u}$
	\EndFor
	\Return $\VEC{f}$
\EndProcedure
\end{algorithmic}
\end{algorithm}

\begin{algorithm}[ht]
\caption{Optimized Recursive Convolution}\label{ALG:RECUR_CONV:OPT}
\begin{algorithmic}[1]
\Procedure{OptRecurConv}{$m$, $\VEC{h}$,  $\VEC{g}$} 
   \INPUT $m$ \qquad \qquad \qquad \quad \   Trim index
   \INPUT $\VEC{h} \in \field{R}^{M}$ \quad \qquad \qquad Fixed FIR
   \INPUT $\VEC{g} \in \field{R}^{N} $ \, \, \,  \qquad  \qquad Exponentiating FIR
   \OUTPUT $\VEC{f} \in \field{R}^{M \plus m(N \minus 1) }$  \quad  \ \ Accumulating FIR
   \State $\underline{m} \gets \FLOOR{\frac{m}{2}}$
   \State $\overline{\underline{m}} \gets M + \underline{m} (N - 1)$
      \vspace{0.25em}
   \If{$m \leq M_{min}$} \Comment{$M_{min}$ is nominally small}
 		\State $\VEC{f} \gets $ DirRecurConv($m$, $\VEC{h}$, $\VEC{g}$)  \Comment{via Algo. \ref{ALG:RECUR_CONV:DIRECT}}
	\Else
			\State $\overline{\VEC{f}} \gets $ $h\RANGEW{\underline{m} + 1}{M} \, \circledast \,  \VEC{g}(\underline{m}) $
   		\State $\underline{\VEC{f}} \gets $ OptRecurConv($h\RANGEW{1}{\underline{m}}$, $\VEC{g}$)   
   		\State $\VEC{u} \gets \BRAK{\underline{\VEC{f}}, \, \VEC{0}^{ M \minus \underline{m} } } + \BRAK{\VEC{0}^{\underline{m}}, \,  \overline{\VEC{f}} \, }$ 
   		   \vspace{0.25em}
   		   \State $\VEC{v} \gets \textrm{OptRecurConv}(m - \underline{m}, u \RANGEW{\underline{m} + 1 }{ \overline{\underline{m}} }    ) $
   		\State $\VEC{f} \gets \BRAK{ u  \RANGEW{1}{\underline{m}}, \, \VEC{v} }$ 
	\EndIf   
   	\Return $\VEC{f}$
\EndProcedure
\end{algorithmic}
\end{algorithm}

The asymptotic run-time costs $\BIGO{M^2 N }$ of the direct Algorithm \ref{ALG:RECUR_CONV:DIRECT} are as prohibitive as the direct exponentiating convolution in Algorithm \ref{ALG:EXP_CONV:DIRECT}, and produce similar outputs for minimum-phase $\VEC{g}$.
Fortunately, we derive an analogous but more expensive partitioned formulation to Algorithm \ref{ALG:EXP_CONV:OPT} in Algorithm \ref{ALG:RECUR_CONV:OPT}.
The crucial difference lies in its two partition sizes, $\underline{m}$ and $M + \underline{m}(N-2)$, which do not remain proportional across sub-problems as the latter depends on the sizes of both inputs $\VEC{h}$ and $\VEC{g}$. The work per recursive call is the convolution operation, with computational costs $\BIGO{MN \log (MN) }$ via FFT; approximate max partition depth is $\log_2 M$. The total asymptotic costs of the largest partitioned trace is $\BIGO{MN^2 \PAREN{ \PAREN{ \log (M) }^2 + \log(M) \log(N) }}$.


%

\begin{figure}[htb]
  \centering
\includegraphics[width=0.9\linewidth]{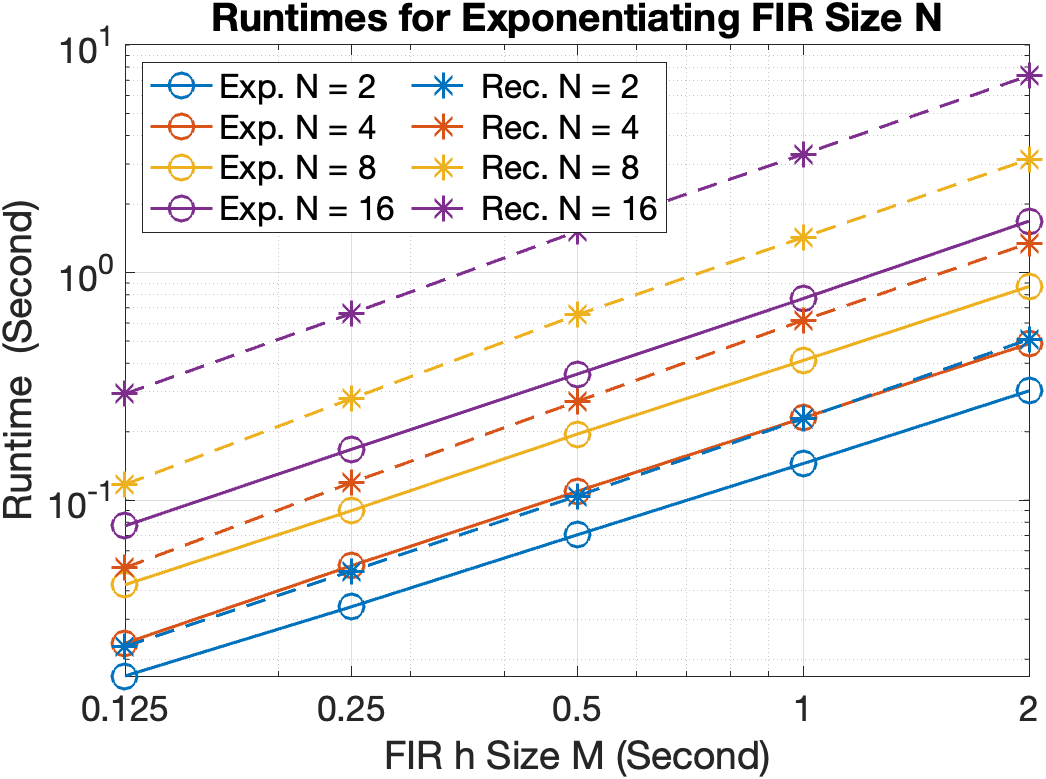} 
\caption{Runtimes (Matlab, Mac M1, $20$ run averages) for exponentiating and recursive Algorithms \ref{ALG:EXP_CONV:OPT}, \ref{ALG:RECUR_CONV:OPT} of varying FIR sizes at $48$ kHz sampling rate}
 \label{FIG:EXP_REC_RUNTIME}
\end{figure}

Lastly, we compare runtime performances of the optimized exponentiating and recursive auto-regressive convolution methods for increasing FIR sizes of $\VEC{h}$, and $\VEC{g}$ in Fig. \ref{FIG:EXP_REC_RUNTIME}. The time-invariant FIR $\VEC{h}$ is sampled from a normal distribution, and the exponentiating filter $\VEC{g}$ is fitted to a randomized spectral-decay target. Both methods exhibit log-linear runtimes w.r.t. size $M$ of the input FIR $\VEC{h}$, but only the exponentiating convolution has log-linear runtimes w.r.t. size $N$ of input filters $\VEC{g}$.

\section{Directional Extensions} 
\label{SEC:DIR}

We now consider extending time-varying exponentiating convolution $f(\VEC{h}, \VEC{g})$ to SH expansions of multi-mic pressure fields. In the general case of multichannel IRs, let us apply $f(\VEC{h}_i, \VEC{g})$ to the $i^{th}$ channel $\VEC{h}_i^T$ transposed along the $i^{th}$ row of the multichannel row-matrix $\MAT{H}$:
\begin{equation} \label{EQ:MULTICH}
\displaystyle
\begin{split}
\MAT{H} = \BRAK{\VEC{h}_1, \hdots, \VEC{h}_{N_H}}^T \in \field{R}^{N_H \times M}, \quad
 \VEC{h}_i \in \field{R}^{M \times 1}. 
\end{split} 
\end{equation} 
Note that multichannel matrices can be linearly transformed under change of bases such as the real or complex SHs, and stored in encoded formats (e.g. Ambisonics B-format). Therefore, it is useful to show that exponentiating convolutions $f(\VEC{h}, \VEC{g})$ commute under all linear transformations $A: \field{C}^{N_H} \rightarrow \field{C}^{N_A}$ given by
\begin{equation} \label{EQ:COMMUTE}
\displaystyle
\begin{split}
A \PAREN{ \BRAK{ \begin{array}{c} f(\VEC{h}_1, \VEC{g}) \\ \vdots \\  f(\VEC{h}_{N_H}, \VEC{g})  \end{array} } } 
& = \BRAK{ \begin{array}{c} f(\tilde{\VEC{h}}_1, \VEC{g}) \\ \vdots \\  f(\tilde{\VEC{h}}_{N_H}, \VEC{g})  \end{array} } \in \field{C}^{N_A \times MN},
\raisetag{2.5ex} 
\end{split} 
\end{equation}
where $A (\MAT{H})   = \BRAK{ \tilde{\VEC{h}}_1 ,  \hdots ,  \tilde{\VEC{h}}_{N_H} }^T$ is the linear transformation of the multichannel matrix $\MAT{H}$. The proof and derivation is given in Appendix Eqs. \eqref{EQ:COMMUTE_EXP_POST}, \eqref{EQ:COMMUTE_EXP_PRE}, which follow from the observation that $A$ linearly transforms only the time-invariant term $\VEC{h}$ in both commutations. Thus, we can directly call Algorithms \ref{ALG:EXP_CONV:OPT}, \ref{ALG:RECUR_CONV:OPT} on any pressure field encoded channels $\tilde{\VEC{h}}$, and recover their time-varying convolved pressure responses by decoding the channels under any linear transformation. Let us now define directional filtering in terms of a linear transformation of channels encoded along SH bases.

\textbf{Directional Filtering:} 
Suppose an instantaneous pressure field function $p_C(r, \theta, \phi) \approx \VEC{Y}(\theta, \phi) \, \VEC{C}(r)$ over time samples $r$ and spherical coordinates $(\theta, \, \phi$), and a time-invariant directional attenuation function $D(\theta, \phi) \approx \VEC{Y}(\theta, \phi)  \, \VEC{D}$ are expanded over finite-order SH bases:
\begin{equation} \label{EQ:SH_DEF}
\displaystyle
\begin{split}
\VEC{Y}(\theta, \phi) & = \BRAK{ Y_0^{0}, \, Y_1^{\minus 1}, \, Y_1^{0}, \, Y_1^{1}, \,  \hdots,  Y_{L_*}^{L_*}  } \in \field{C}^{1 \times (L_* \plus 1)^2 },  \\
 Y_l^m (\theta, \phi) & = \sqrt{\frac{(2l + 1)}{4 \pi} \frac{(l-m)!}{(l+m)!} } \, P_l^m (\cos \theta) e^{i m \phi},
\end{split} 
\raisetag{5ex}
\end{equation}
where $\theta$, $\phi$ are co-latitude and azimuth respectively, and the associated Legendre polynomials $P_l^m(\cos \theta) $ are indexed via expansion order $m$ bounded by $-l \leq m \leq l$, and degree $l$ bounded by $0 \leq l \leq L_*$ for max-degree $L_*$. The expansion weights, given by the column matrix $\MAT{C} = \BRAK{\VEC{C}(1), \hdots ,  \VEC{C}(M) } \in \field{C}^{N_C \times M}$ for $N_C = (L_C + 1)^2$, and $\VEC{D} \in \field{C}^{N_D \times 1}$ for $N_D = (L_D + 1)^2$, are projections of the pressure and attenuation functions on the SH bases respectively. 

Let us define a directional weighted pressure field $p_E (r, \theta, \phi)$ by the following product of attenuation and pressure field functions under SH expansions:
\begin{equation} \label{EQ:DIR_FILTER_MODEL}
\displaystyle
\begin{split}
p_E (r, \theta, \phi) =  D (\theta, \phi) \,  p_C (r, \theta, \phi)  \approx \VEC{Y}(\theta, \phi) \, \VEC{E}, 
\end{split} 
\end{equation}
where its SH expansion can be expressed in terms of $\VEC{E} \in \field{C}^{N_E \times 1}$ weighted SH bases for $N_E = (L_D + L_C + 1)^2$ as the SH product operator has closure under addition \cite{pfaff2017filtering, luo2021spherical}. 
Furthermore, the SH product operator can be decomposed into a linear transform $A(\VEC{D})$ of the directional filter coefficients $\MAT{D}$ represented by the \textit{transfer matrix} \cite{jarosz08thesis} $\MAT{A}  \in \field{C}^{N_E \times (L_C \plus 1 )^2}$ whose entrants are given in Appendix Eq. \eqref{EQ:SH_PROD_MAT_A}. Thus, the weighted expansion coefficients  $\VEC{E} = \MAT{A} \VEC{C}$ simplify into a matrix-vector product of the pressure field's expansion coefficients.

\begin{figure*}[h]
\centering 
    \subfloat[Direction attenuation function $D(\theta, \phi)$ \label{FIG:SAMPLE_DIR_FILTER}]{%
        \includegraphics[height=0.1325\textheight]{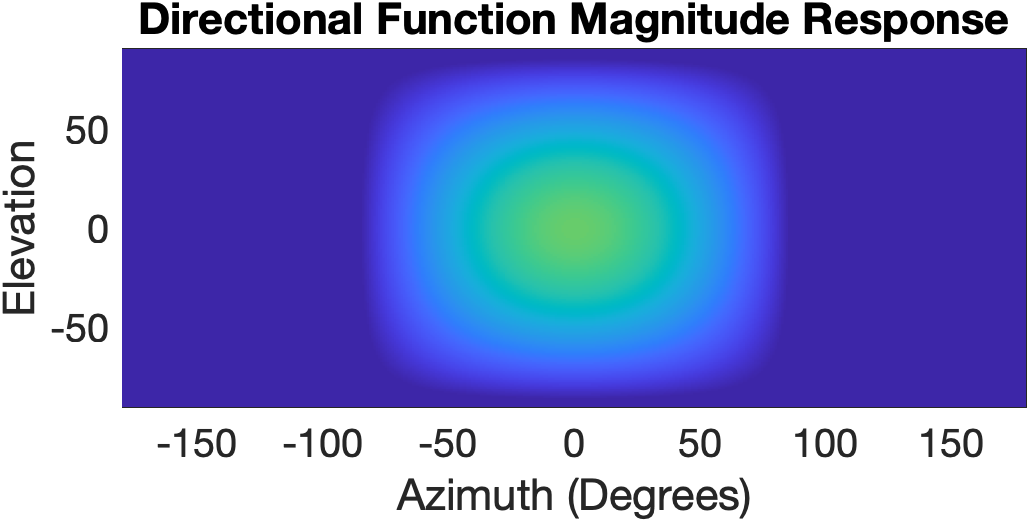}%
    }\hspace{1pt}
    \subfloat[Pressure field function $p_C(r, \theta, \phi)$  \label{FIG:SAMPLE_RAND_FUNC}]{%
        \includegraphics[height=0.1325\textheight]{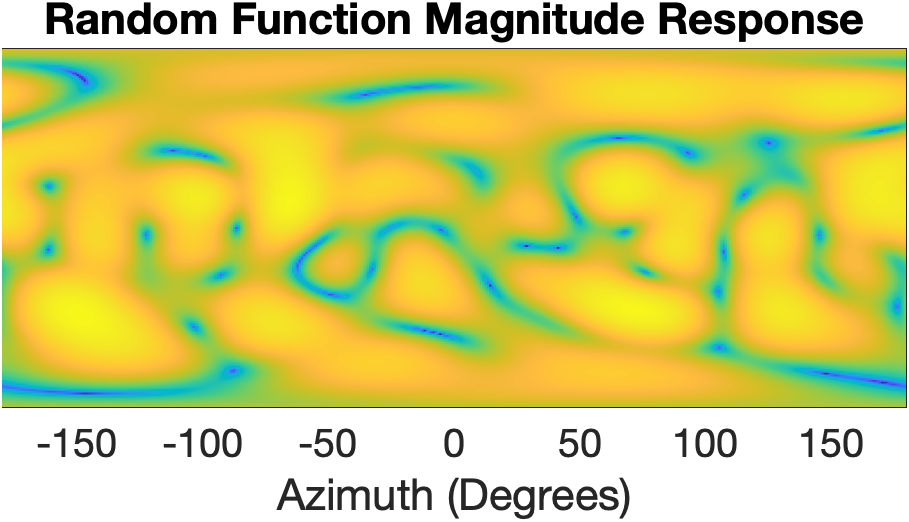}%
    }\hspace{1pt}
        \subfloat[Weighted pressure field $p_E(r, \theta, \phi)$  \label{FIG:SAMPLE_FILTERED_RAND_FUNC}]{%
        \includegraphics[height=0.1325\textheight]{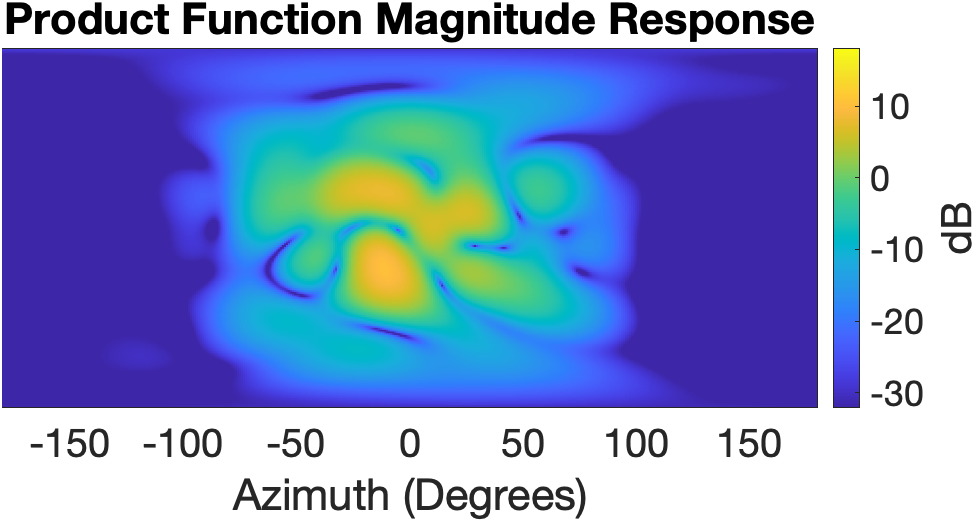}%
    }
      \caption{\label{FIG:SAMPLE_RANDOM_DIR_FILTER}We expand the kernel function of chordal distances \cite{luo2021spherical} given by $k(d) = e^{\minus \ABS{d}^2 / (2 \ell) }$ at the expansion center $\PAREN{\theta_c = 90^\circ, \, \phi_c = 0^\circ}$ upto max-degree $L_D = 7$ in Fig. \ref{FIG:SAMPLE_DIR_FILTER}, where $d$ is the Euclidean distance between spherical coordinates $(\theta, \phi)$ and $(\theta_c, \phi_c)$ on the unit sphere . The product of the directional function and randomized pressure field in Fig. \ref{FIG:SAMPLE_RAND_FUNC} (max-degree $L_C = 7$) attenuates the directional responses in Fig. \ref{FIG:SAMPLE_FILTERED_RAND_FUNC}.}    
\end{figure*}

In practice, the product function and SH weights in Eq. \eqref{EQ:DIR_FILTER_MODEL} can be efficiently computed via forward and inverse fast spherical harmonic transform (FSHT) \cite{suda2002fast}:
\begin{equation} \label{EQ:DIR_FILTER_MODEL_SHT}
\displaystyle
\begin{split}
 \VEC{E} & =  \MAT{Y}^{\minus 1}(\VEC{\theta}, \VEC{\phi})  \PAREN{ \MAT{Y}_D(\VEC{\theta}, \VEC{\phi})  \, \VEC{D}   \, \odot \, \MAT{Y}_{C}(\VEC{\theta}, \VEC{\phi} ) \, \VEC{C}(r)  }   \\
 & = \textrm{FSHT} \PAREN{ \textrm{IFSHT}(\VEC{D})   \, \odot \, \textrm{IFSHT}( \VEC{C}(r) ) } \\
 & \approx \textrm{FSHT} \PAREN{ D(\VEC{\theta}, \VEC{\phi})   \, \odot \, \textrm{IFSHT}( \VEC{C}(r) ) }, 
\end{split} 
\end{equation}
where $\odot$ is element-wise multiplication, and $\VEC{\theta}, \VEC{\phi} \in \field{R}^{N_E}$ are spherical coordinates of a uniform and high-resolution grid sampled over a sphere. Therefore, the square matrix $\MAT{Y}(\VEC{\theta}, \VEC{\phi} ) \in \field{C}^{N_E \times N_E }$ of SH evaluations at the grid coordinates is  well-conditioned and invertible. The rectangular matrices $\MAT{Y}_D(\VEC{\theta}, \VEC{\phi} )$, $\MAT{Y}_C(\VEC{\theta}, \VEC{\phi} )$ truncate $\MAT{Y}(\VEC{\theta}, \VEC{\phi} )$ to the first $N_D$, $N_C$ columns respectively. For illustration, we specify a directional attenuation function by expanding a squared exponential kernel function over the spherical coordinates, and whose product with a randomized pressure field is shown in  Fig. \ref{FIG:SAMPLE_RANDOM_DIR_FILTER}.

We now extend time-varying convolutions to SH bases for the case of direction independent filters $\VEC{g}$. The equivalences in Eq. \eqref{EQ:DIR_FILTER_MODEL_SHT} and linearity of Eq. \eqref{EQ:COMMUTE} permits us to compute either $f(\MAT{C}, \, \VEC{g})$ directly on each of the $N_C$ rows of the pressure field's SH expansion coefficients $\MAT{C}$, or alternatively $f \PAREN{ \MAT{P},  \, \VEC{g} }$ on each of the $N_E$ rows on the spherical grid coordinates of the pressure field matrix $\MAT{P} = \MAT{Y}_C (\VEC{\theta}, \VEC{\phi})  \, \VEC{C} \in \field{R}^{N_E \times M}$.
Therefore, the following time-varying convolved SH expansion matrix $\tilde{\MAT{E}} \in \field{C}^{N_E \times MN}$ has two equivalences:
\begin{equation} \label{EQ:DIR_FILTER_MODEL_SHT_EXP_CONV}
\displaystyle
\begin{split}
\tilde{\MAT{C}} & =  \BRAK{ \begin{array}{c} \tilde{\VEC{C}}_1 \\ \vdots \\ \tilde{\VEC{C}}_{(L_C \plus 1)^2}  \end{array}  },   \quad
\tilde{\VEC{C}}_i =  f(\BRAK{C_{i1}, \hdots, C_{iM} }, \, \VEC{g}), \\[2pt]
\tilde{\MAT{P}} & =  \BRAK{ \begin{array}{c} \tilde{\VEC{P}}_1 \\ \vdots \\ \tilde{\VEC{P}}_{(L_D \plus L_C \plus 1)^2} \end{array}  } ,  \quad
\tilde{\VEC{P}}_i =  f(\BRAK{P_{i1}, \hdots, P_{iM} }, \, \VEC{g}), \\[3pt]
 \tilde{\MAT{E}} & =  \MAT{Y}^{\minus 1}(\VEC{\theta}, \VEC{\phi})  \PAREN{ \MAT{Y}_D (\VEC{\theta}, \VEC{\phi})  \, \VEC{D} \VEC{1}^T   \, \odot \, \MAT{Y}_C (\VEC{\theta}, \VEC{\phi} ) \, \tilde{\VEC{C}}  } , \\
  & =  \MAT{Y}^{\minus 1}(\VEC{\theta}, \VEC{\phi})  \PAREN{ \MAT{Y}_D (\VEC{\theta}, \VEC{\phi})  \, \VEC{D} \VEC{1}^T   \, \odot \, \tilde{\MAT{P}}  } , \\
\end{split} 
\raisetag{2.5ex}
\end{equation}
where computing over the SH expansion coefficients $\tilde{\MAT{C}}$ requires $L_D^2 + 2 L_D(L_C + 1)$ fewer evaluations of $f$ than computing over the sample pressure field matrix $\MAT{P}$. Next, we consider the case of time-varying filtering with direction dependent filters $\VEC{g}(\theta, \phi)$.

\textbf{Directional Reverberation:} 
Suppose we allow exponentiating filters $\VEC{g}$ for pressure field $\tilde{\VEC{P}}$ in Eq. \eqref{EQ:DIR_FILTER_MODEL_SHT_EXP_CONV} to vary as if sampled over spherical coordinates via $\VEC{g}(\theta, \phi)$. The rows of $\tilde{\VEC{P}_i}$ are pressure time-series on the spherical grid coordinates $(\theta_i, \phi_i) \in (\VEC{\theta}, \VEC{\phi})$ become
\begin{equation} \label{EQ:DIR_REVERB_P}
\displaystyle
\begin{split}
\tilde{\VEC{P}}_i =  f(\BRAK{P_{i1}, \hdots, P_{iM} }, \, \VEC{g}(\theta_i, \phi_i) ),
\end{split} 
\end{equation}
whereby rows of pressure matrix $\MAT{P}$ at the grid coordinates undergo direction dependent convolution.
In the case of unit sized  $(N=1)$  filters $\VEC{g}(\theta, \phi)$, the time-varying convolution process is equivalent to exponential-decaying directional filtering of Eq. \eqref{EQ:DIR_FILTER_MODEL}, where the direction attenuation function $D(\theta, \phi) = \VEC{g}^m(\theta, \phi)$ at time-sample $m$ exponentiates a single coefficient. For higher-order filters, we require the filter's frequency response $\left . G(\omega, \theta, \phi) = G(z, \theta, \phi) \right | z = e^{j \omega}$ for angular frequency $\omega$, to smoothly vary over spherical coordinates as to recover a low-order expansion matrix $\tilde{\MAT{E}}$.
Consider the gradient of the exponentiated filter's magnitude frequency responses $\ABS{G^m(\omega, \theta, \phi)}$ w.r.t. spherical coordinates given by
\begin{equation} \label{EQ:DIFF_MAG_EXP}
\displaystyle
\begin{split}
\nabla \ABS{G^m (\omega, \theta, \phi)} & = m \ABS{G(\omega, \theta, \phi)}^{m \minus 1}  \nabla \ABS{G (\omega, \theta, \phi)}.
\end{split} 
\end{equation}
The gradient $\nabla \ABS{ G (\omega, \theta, \phi)}$ should be differentiable for non-zero magnitudes, and the exponentiating term grows unbounded unless we restrict the filter's magnitude responses $0 < \ABS{G(\omega, \theta, \phi)} \leq \tau < 1$. Thus, we model $\ABS{G(\omega, \theta, \phi)}$ as follows:

Let the magnitude response $\ABS{G(\omega, \theta, \phi)}$ achieve a target reverberation time of $T_{60} (\omega, \theta, \phi)$ seconds for $60$ decibel (dB) attenuation.  Equating the exponentiated magnitude response $\ABS{G^m (\omega, \theta, \phi)} = 10^{\minus 3}$ for the attenuation's sample time $m = F_s \, T_{60} (\omega, \theta, \phi)$  at the sampling rate $F_s$ yields the following magnitude dB target:
\begin{equation} \label{EQ:RT60_MAG}
\displaystyle
\begin{split}
\ABS{G(\omega, \theta, \phi)}_{dB} = \frac{\minus 60}{F_s \, \textrm{T}_{60} (\omega, \theta, \phi)  },
\end{split} 
\end{equation}
by which substituting into the exponential term $m \ABS{G(\omega, \theta, \phi)}^{m \minus 1} $ in Eq. \eqref{EQ:DIFF_MAG_EXP} gives the following upper bound on the exponential magnitude component:
\begin{equation} \label{EQ:DIFF_MAG_EXP_T60}
\displaystyle
\begin{split}
m_*   = \argmax{m} m \ABS{G(\omega, \theta, \phi)}^{m \minus 1}  =  \frac{F_s \, T_{60} (\omega, \theta, \phi)}{3 \log 10 }, \\
   \Rightarrow m \ABS{G(\omega, \theta, \phi)}^{m \minus 1} \leq   10^{\frac{3}{F_s T_{60}(\omega, \theta, \phi) }}  \frac{ F_s \, T_{60}(\omega, \theta, \phi) }{ (3 \log 10) e }, 
\end{split} 
\end{equation}
which is approximately linear w.r.t. $T_{60}(\omega, \theta, \phi)$. Thus, longer reverberation times induce larger gradients in Eq. \eqref{EQ:DIFF_MAG_EXP} that must be offset by smaller Euclidean norms of the gradient component $\NORM{ \nabla \ABS{ G (\omega, \theta, \phi)} }$ such that larger $T_{60}(\omega, \theta, \phi)$ are smoother over spherical coordinates.
Note that in practice, we observe longer reverberation times at lower frequencies in RIRs, and model distributions of $T_{60}(\omega, \theta, \phi)$ whose smoothness in the spherical coordinates vary inversely with frequency. We present a modified non-stationary GP \cite{paciorek2003nonstationary} for modeling $T_{60}(\omega, \theta, \phi)$ distributions in the next section.

With regards to smooth phase components of the exponentiated filter's transfer function $G^m(\omega, \theta, \phi)$, the group delay of $G(\omega, \theta, \phi)$ should be minimized as the exponentiation operation multiplies the phase response $\angle G^m(\omega, \theta, \phi) = m \angle G(\omega, \theta, \phi)$. We therefore specify the filter's response to be near minimum-phase when subjected to constraints on its upper-bound magnitude $\ABS{G(\omega , \theta, \phi)} \leq \tau $, which ensures attenuation across all frequencies and decreases the gradient in Eq. \eqref{EQ:DIFF_MAG_EXP}. The minimum-phase response following the real-cepstrum method \cite{oppenheim1999discrete,pei2006minimum} can be derived from the desired magnitude responses of Eq. \eqref{EQ:RT60_MAG} at $\overline{N}$ number of uniformly spaced frequencies between DC and $F_s$. We enforce the latter magnitude bounds via quadratic constrained least-squares fitting of the $N$-tap FIR filter $\VEC{g}$ to $\overline{N}$ number of minimum-phase frequency response targets.

Consider the following second-order cone program:
\begin{equation} \label{EQ:NEAR_MIN_PHASE}
\displaystyle
\begin{split}
& (\lambda_*, \VEC{g}_*)   = \argmin{(\lambda, \VEC{g})} \lambda, \quad \textrm{s.t. } \quad  
 \NORM{  \MAT{W}  \VEC{g}- \VEC{b} } \leq \lambda, \quad \lambda \geq 0, \\ 
& \NORM{  \VEC{w}^T(k)  \,  \VEC{g}  } \leq \tau, \quad  \forall k \in \CBRAK{0, \hdots, \FLOOR{\frac{K}{2}}  }, \quad \textrm{Constraints} \\[3pt]
& w_n(k) = e^{\minus j 2 \pi k (n \minus 1) / K }, \quad
W_{kn} =  e^{\minus j  2 \pi (k \minus 1) (n \minus 1) / \overline{N} }, 
\end{split}
\raisetag{3ex}
\end{equation}
where $\MAT{W} \in \field{C}^{\overline{N} \times N}$ is an $\overline{N}$-sized discrete Fourier transform matrix truncated to the first $N$ columns, and $\VEC{b} = \BRAK{G(z_1), \hdots, G(z_{\overline{N}})}^T \in \field{C}^{\overline{N} \times 1}$ the vector of target minimum-phase responses at frequencies $\VEC{\omega} = \CBRAK{z_1, \hdots, z_{\overline{N}}} $ where $z_n = e^{-j 2 \pi (n - 1) / \overline{N} }$. 
Minimizing the non-negative variable $\lambda$ in the objective function minimizes the error norm between the filter and target responses; 
we can truncate $\MAT{W}$,  $\VEC{b}$ to the first $\FLOOR{\overline{N} / 2} + 1$ rows such that the errors span only the frequencies between DC and Nyquist.
The magnitude responses of $\VEC{g}$ are constrained over $K$ uniform and densely spaced angular frequencies evaluated at the Fourier bases  $\VEC{w}^T(k) = \BRAK{w_1(k), \hdots, w_N(k)} \in \field{C}^{1 \times N} $. Therefore, we can fit fewer coefficients to the number of target responses and constraints as $N \leq \overline{N} \ll K$.

\textbf{Gaussian Process Model:} 
Let us now model a distribution of smooth $T_{60}(\omega, \theta, \phi)$ functions sampled from a GP with model supports given by the following observed logarithmic reverberation time vector $\VEC{y} = \BRAK{y_1, \hdots, y_S}^T \in \field{R}^{S \times 1} $ at the set of frequency and spherical coordinates  $\VEC{X} = \CBRAK{\VEC{x}_1, \hdots, \VEC{x}_S}$, $\VEC{x}_n  = \PAREN{\underline{\omega_n}, \,  \underline{\theta_n},  \, \underline{\phi_n} }$:
\begin{equation} \label{EQ:GPR_SUPPORT}
\displaystyle
\begin{split}
y_n  = \log T_{60}(\underline{\omega_n}, \, \underline{\theta_n},  \, \underline{\phi_n} ) -  \mu(\underline{\omega_n}),
\end{split} 
\end{equation}
where $\VEC{y}$ is centered over a prior log-mean power-law reverberation time function  $\mu(\omega)  = \log \overline{T}_{60}(\omega) $ of ordinary frequency $\nu(\omega)$  given by   
\begin{equation} \label{EQ:GPR_PRIOR_MU}
\displaystyle
\begin{split}
\overline{T}_{60}(\omega) =  \alpha \PAREN{\nu(\omega)}^{- \beta}, \quad
\nu(\omega) =  \frac{\angle \omega}{2 \pi},
\end{split} 
\end{equation}
such that the scale $\alpha$ and power $\beta$ parameters are non-negative. The mean function detrends the log reverberation time over frequency, and the earlier log-transformation ensures that sampled reverberation time functions from the following GP distributions are positive:
We can sample $T_{60}(\omega, \theta, \phi)$ functions over the input set $\VEC{X}_*$ of size $S_*$ from either the GP prior normal distribution $\mathcal{N}(\VEC{\mu}(\VEC{X}_*), \,  \VEC{k}(\VEC{X}_*, \VEC{X}_*) )$  or the predictive normal distribution $\mathcal{N}(\VEC{\mu}_* (\VEC{X}_*), \,  \MAT{\Sigma}_* (\VEC{X}_* )  )$ conditioned on observed $\VEC{X}$, $\VEC{y}$. The means and covariances follow
\begin{equation} \label{EQ:GPR}
\displaystyle
\begin{split}
 \VEC{\mu} (\VEC{X}_* ) & = [\mu(\omega_1), \hdots,  \mu (\omega_{S_*}) ]^T, \,\,\,\, \,\, \qquad \textrm{Prior} \\
\VEC{\mu}_* (\VEC{X}_*)  & = \VEC{\mu} (\VEC{X}_* ) + \VEC{k}(\VEC{X}_*, \VEC{X})  \, \underline{\MAT{K}}^{\minus 1} \VEC{y}, \quad \textrm{Posterior} \\
\MAT{\Sigma}_* (\VEC{X}_* ) & =  \VEC{k}(\VEC{X}_*, \VEC{X}_*) -  \VEC{k}(\VEC{X}_*, \VEC{X})   \,  \underline{\MAT{K}}^{\minus 1}  \, \VEC{k}(\VEC{X}, \VEC{X}_*) ,\\
\underline{\MAT{K}} & = \MAT{k}\PAREN{\VEC{X}, \VEC{X}} + \MAT{\Lambda} \in \field{R}^{S \times S},
\end{split} 
\end{equation}
where entrants of $\MAT{k}(\VEC{X}, \VEC{X}_*) \in \field{R}^{S \times S_*}$ are covariance functions $k_{ij} = k(\VEC{x}_i, \VEC{x}_j)$ of tuples in $\VEC{X}$, $\VEC{X}_*$, and $ \MAT{\Lambda}$ is the diagonal matrix of noise variances. 

The covariance function $k(\VEC{x}, \VEC{x}')$ determines the smoothness of functions drawn from a GP w.r.t. similarity between the inputs $\VEC{x}$ and $\VEC{x}'$. We consider the following covariance function that is stationary in the spherical coordinates $(\theta, \phi)$, and non-stationary in wavelengths $\lambda = \ell / (\nu(\omega))^{\gamma}$ where $\ell$, $\gamma$, $\sigma$ are velocity, power, and scaling hyper-parameters respectively:
\begin{equation} \label{EQ:GPR_COV}
\displaystyle
\begin{split}
&   k(\VEC{x}, \VEC{x}')  =
\sigma^2   \PAREN{\frac{2 \, \lambda \lambda'}{ \lambda^2 + \lambda'^2   } }^{\frac{3}{2}} 
  \exp \PAREN{- \frac{ d^2(\theta, \phi, \theta', \phi') }{  \PAREN{\lambda^2 + \lambda'^{2} } / \, 2  } }, \\
&   d^2(\theta, \phi, \theta', \phi') = 2 \sin \PAREN{ \frac{ \ABS{\cos^{\minus 1} \PAREN{\VEC{v}^T \VEC{v}'}  } }{2} } 
= \NORM{\VEC{v} - \VEC{v}'}_2, 
\end{split} 
\raisetag{17ex}
\end{equation}
where $\VEC{v}, \VEC{v}' \in \field{R}^{3\times 1} $ the Cartesian unit-directions of the spherical coordinates $(\theta, \phi)$, $(\theta', \phi')$ respectively, and $d$ is the chordal distance between the points on a unit-sphere. Note that $ k(\VEC{x}, \VEC{x}')$ is a valid non-stationary squared exponential covariance function for the $3$-dimensional case in \cite{paciorek2003nonstationary}. The non-stationarity in wavelengths induces a slower decaying exponential over the chordal distance for larger $\lambda$. As a result, reverberation times sampled at longer wavelengths in lower frequencies are smoother over the spherical coordinates.

\section{Experiments}
\label{SEC:EXP}

\begin{figure*}[h]
\centering 
    \subfloat[Sample $T_{60}$ functions drawn from the GP prior at a common spherical coordinate \label{FIG:GP_PRIOR}]{%
        \includegraphics[width=0.32\textwidth]{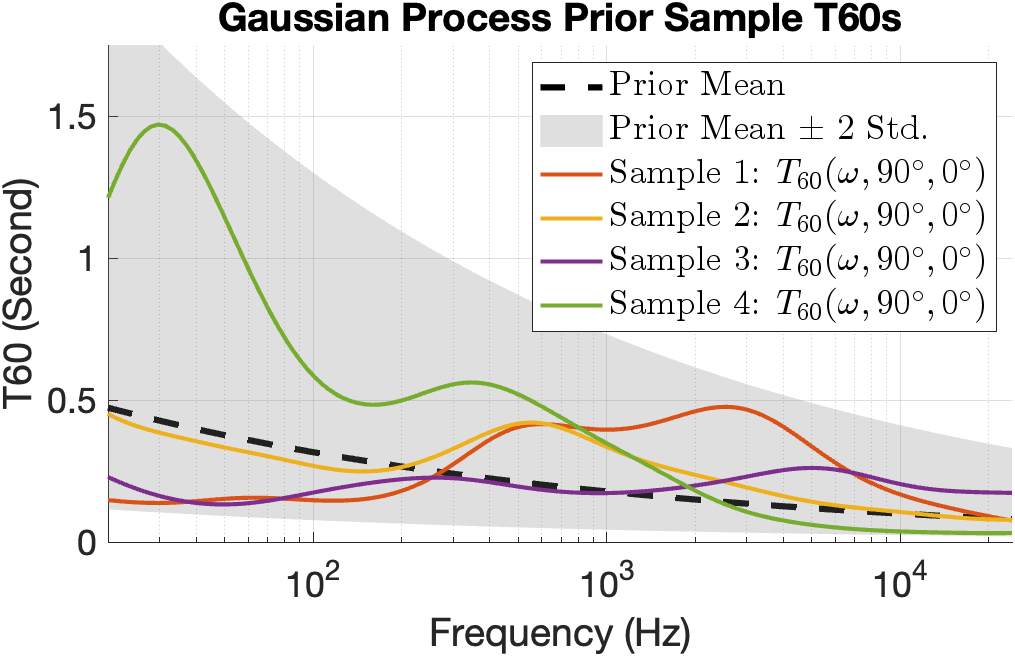}%
    }\hspace{1pt}
            \subfloat[Sample $T_{60}$ functions drawn from the GP posterior conditioned on $4$ observations \label{FIG:GP_POST}]{%
        \includegraphics[width=0.32\textwidth]{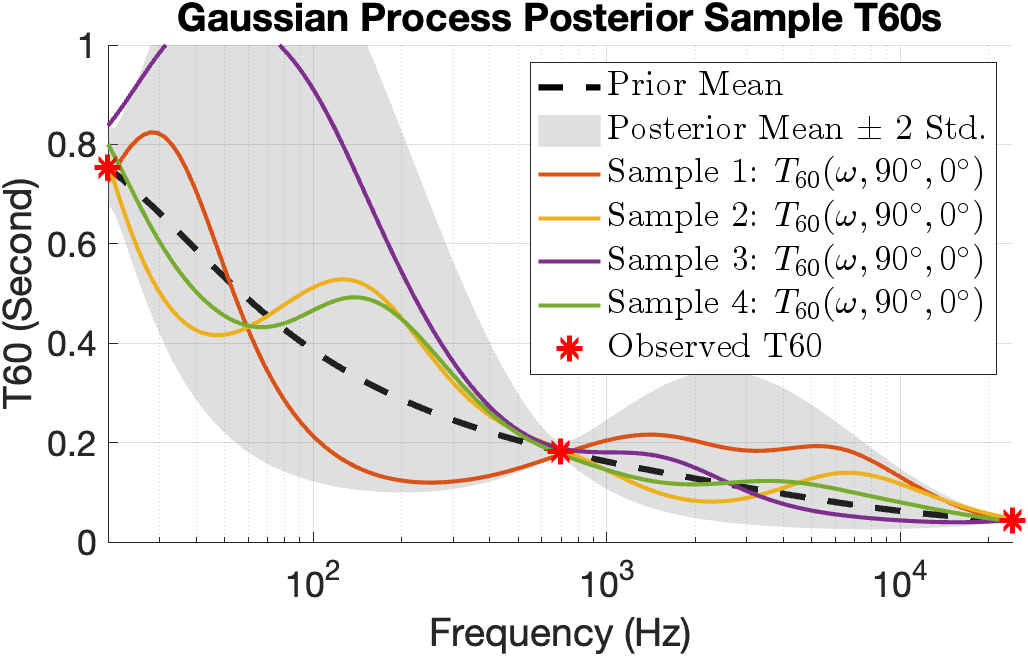}%
    }\hspace{1pt}
    \subfloat[Evaluation grid's $T_{60}$ functions drawn from the GP prior on the horizontal plane  \label{FIG:GP_PRIOR_GRID}]{%
        \includegraphics[width=0.32\textwidth]{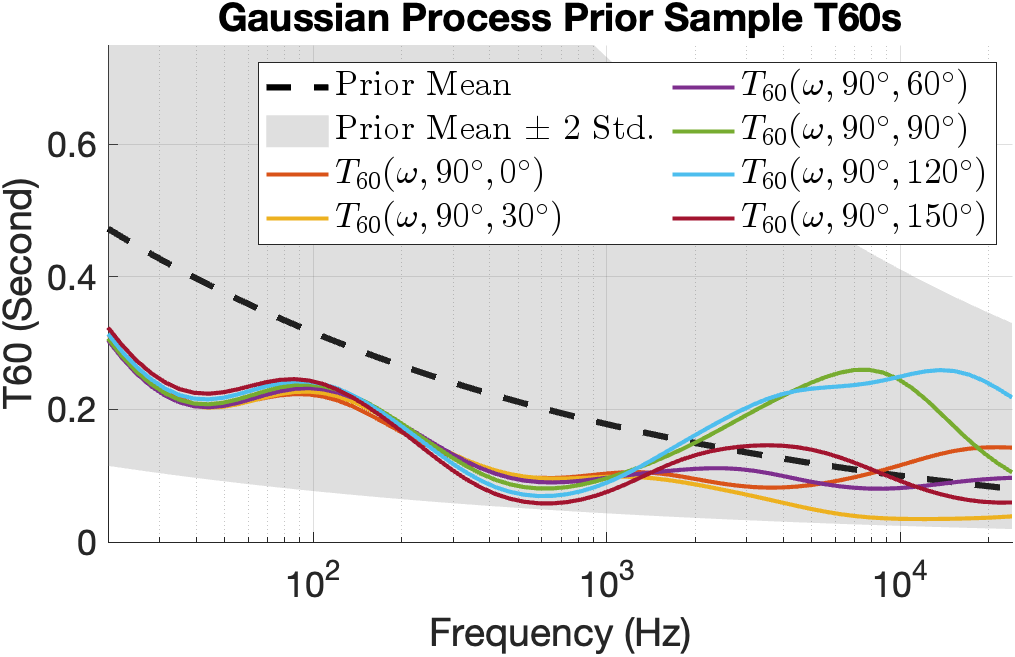}%
    }
      \caption{\label{FIG:GP_DIST}Sample $T_{60}$ functions that are independently drawn from the GP prior in Fig. \ref{FIG:GP_PRIOR} vary over a wide range in time/frequency, and narrows when drawn from the GP posterior predictive distribution in Fig. \ref{FIG:GP_POST}. The $T_{60}$ functions jointly sampled over various spherical coordinates on the same grid in Fig. \ref{FIG:GP_PRIOR_GRID} remain correlated in lower frequencies, and accords with the design of our non-stationary covariance function in frequency.
      }    
\end{figure*}

\textbf{Gaussian Process T60 Sampling:} 
%
Let the $T_{60}$ sampling grid $\CBRAK{\VEC{\omega} \times  \PAREN{ \VEC{\theta}, \VEC{\phi} } }$ be the Cartesian product between the set of frequencies for fitting $\VEC{g}$ in Eq. \eqref{EQ:NEAR_MIN_PHASE}, and the uniform spherical coordinate grid for recovering SH expansion coefficients $\tilde{\MAT{E}}$ in Eq. \eqref{EQ:DIR_FILTER_MODEL_SHT_EXP_CONV}.
We specify the set of sampling inputs $\VEC{X}_* = \CBRAK{\VEC{x}_{11}, \hdots,  \VEC{x}_{\overline{N} N_E}}$, where  $\VEC{x}_{mn} = \PAREN{\omega_m, \theta_n, \phi_n}$, and draw sample $T_{60}(\omega, \theta, \phi)$ functions from the GP prior and predictive distribution in Eq. \eqref{EQ:GPR} as shown in Fig. \ref{FIG:GP_DIST}.  The power-law function's parameters $\alpha = 1$, $\beta = 0.25$ yield a tapering GP prior mean $\mu(\omega)$ and $\overline{T}_{60}$ in Eq. \eqref{EQ:GPR_PRIOR_MU}, and narrows the variance at higher frequencies due to the log-transformation $\log T_{60}$ in Eq. \eqref{EQ:GPR_SUPPORT} and covariance function in Eq. \eqref{EQ:GPR_COV} ($\sigma = \sqrt{2}/2$, $\gamma = 2/3$, $\ell = 343$). Conditioning the GP on several observed $T_{60}$ values at one spherical coordinate yields a predictive distribution with variances that tighten near observed $\VEC{\omega}$, and revert to the prior when afar. Jointly sampling several $T_{60}$ functions across different spherical coordinates exhibits positive co-varying times at lower frequencies due to larger mean-squared wavelengths $\lambda$ in the non-stationary covariance of Eq. \eqref{EQ:GPR_COV}.

\begin{figure}[ht]
  \centering
\includegraphics[width=0.9\linewidth]{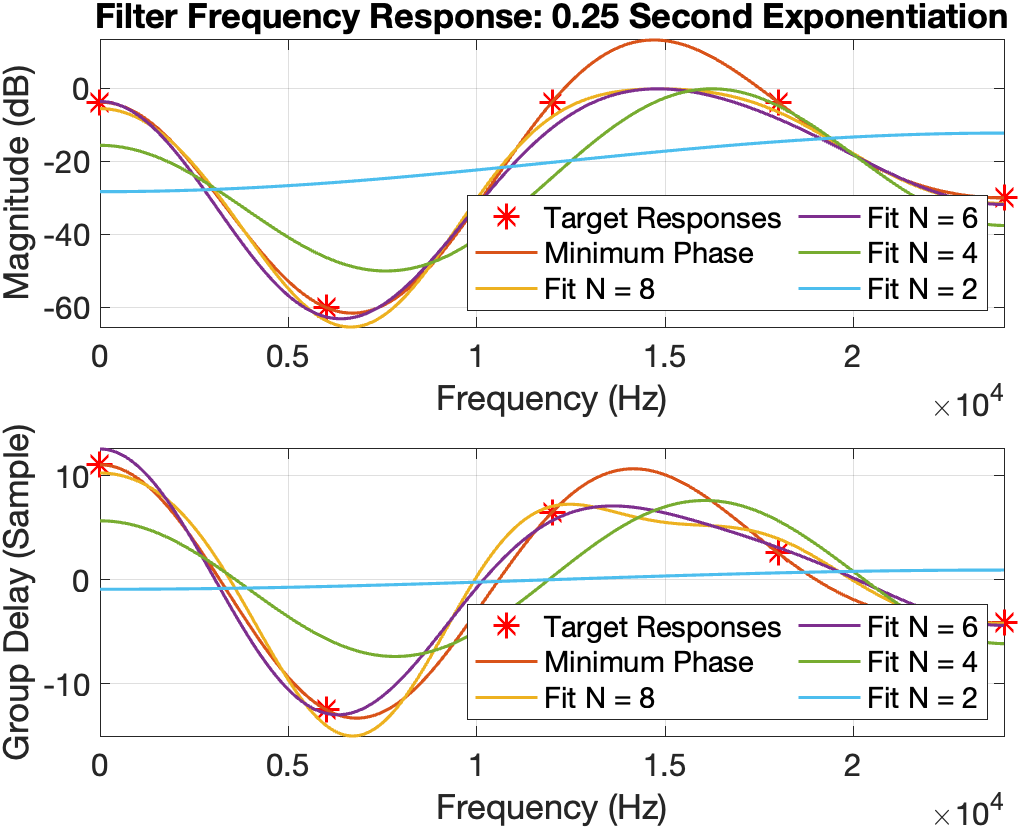} 
\caption{The frequency responses of filters $\VEC{g}$, fitted with fewer $N$ taps to $T_{60}$ targets via Eq. \eqref{EQ:NEAR_MIN_PHASE}, exhibit amplified errors under exponentiation.}
 \label{FIG:FILTER_FIT}
\end{figure}

\textbf{Exponentiating Filter Sensitivity:}  
%
Under exponentiation, we expect the transfer functions $G^m(z)$ of $\VEC{g}$ in Eq. \eqref{EQ:EXP_CONV_TF} to amplify the filter's fitting errors from Eq. \eqref{EQ:NEAR_MIN_PHASE} as magnitude dB and group-delay responses are multiplied under exponentiation. We show that reducing the filter size $N$ far below the number of target $T_{60}$ responses over frequency introduces large errors. For illustration, consider the following $T_{60} = \BRAK{4, 0.25, 4, 4, 0.5}$ second targets at uniform spaced frequencies between DC and Nyquist at 48 kHz sampling rate in Fig. \ref{FIG:FILTER_FIT}. Specifying the target log-magnitude responses following Eq. \eqref{EQ:RT60_MAG} and reflecting over Nyquist yields $N = 8$ tap minimum-phase targets, but which its unconstrained filter responses exceed unity between $12-16$ kHz. The constrained filters' fitted responses for $N \in \CBRAK{8, 6}$ satisfy the unity bounds $\tau = 1$ over the frequency range and have low fitting errors. Further decreasing $N$ results in both significant deviations from intended target responses, and larger variations across fitted responses over the spherical coordinates. We can decrease the error variance by increasing the $T_{60}$ resolution in frequency over the GP evaluation grid.

\begin{figure*}[ht]
\centering 
    \subfloat[$T_{60}(\omega, \theta, \phi)$ at uniform spherical coordinates \label{FIG:SH_EXP_CONV:T60}]{%
        \includegraphics[width=0.34\textwidth]{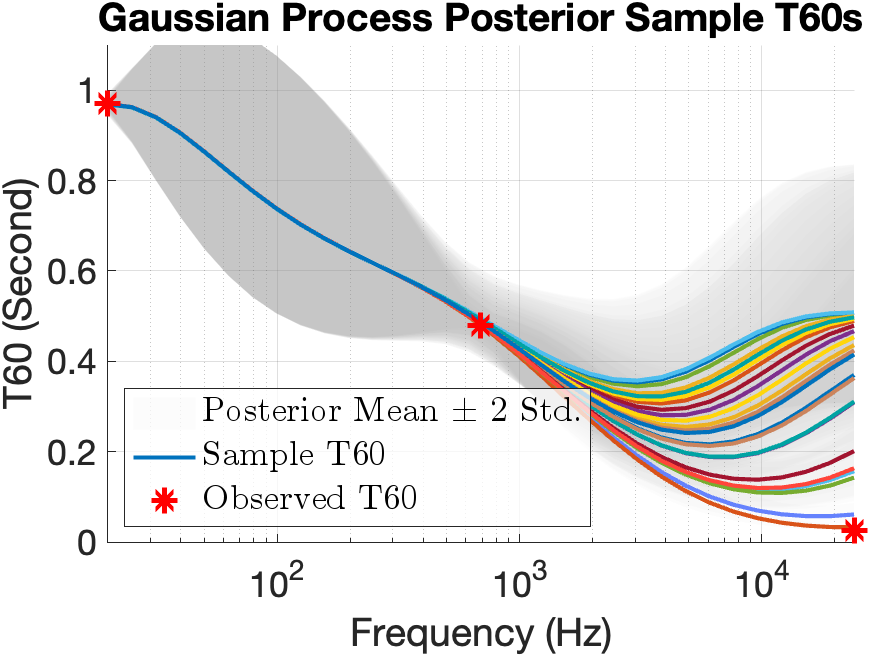}%
    }\hspace{2pt}
    \subfloat[$\phi = 0^\circ$\label{FIG:SH_EXP_CONV:SAMP_1}]{%
        \includegraphics[width=0.1625\textwidth]{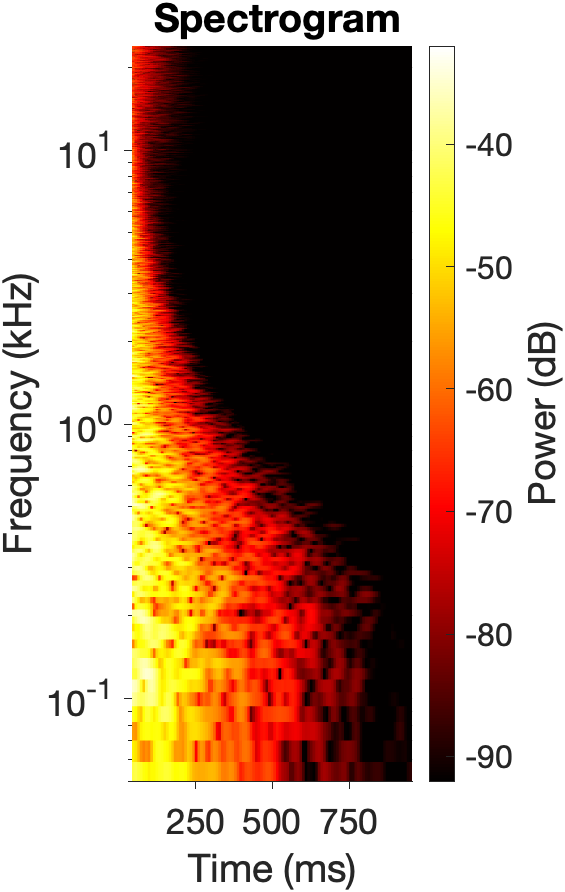}%
    }
    \subfloat[$\phi = 60^\circ$\label{FIG:SH_EXP_CONV:SAMP_2}]{%
        \includegraphics[width=0.1625\textwidth]{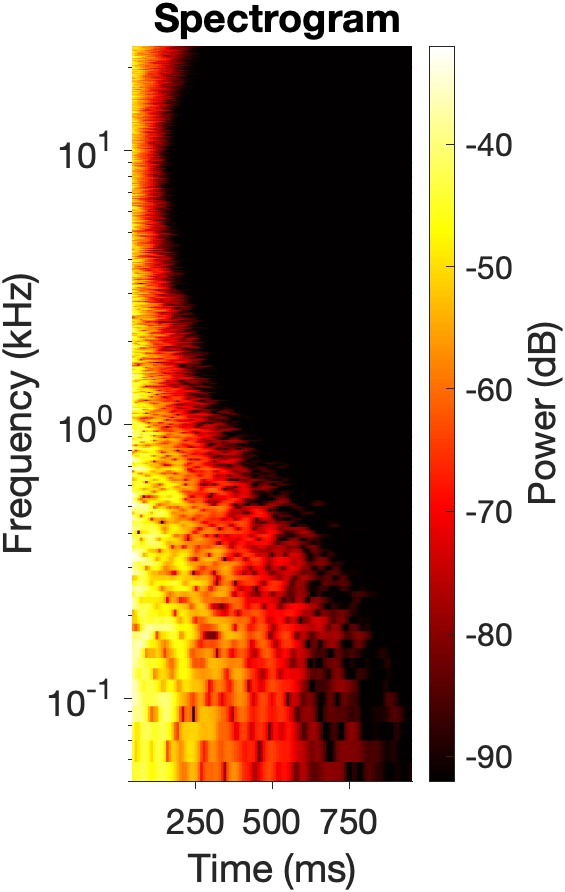}%
    }
    \subfloat[$\phi = 120^\circ$\label{FIG:SH_EXP_CONV:SAMP_3}]{%
        \includegraphics[width=0.1625\textwidth]{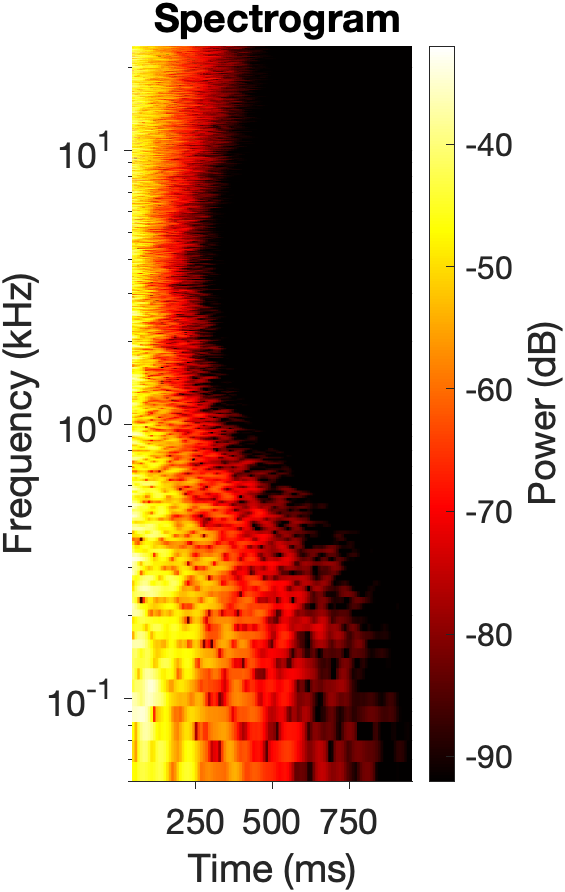}%
    }
     \subfloat[$\phi = 180^\circ$\label{FIG:SH_EXP_CONV:SAMP_4}]{%
        \includegraphics[width=0.1625\textwidth]{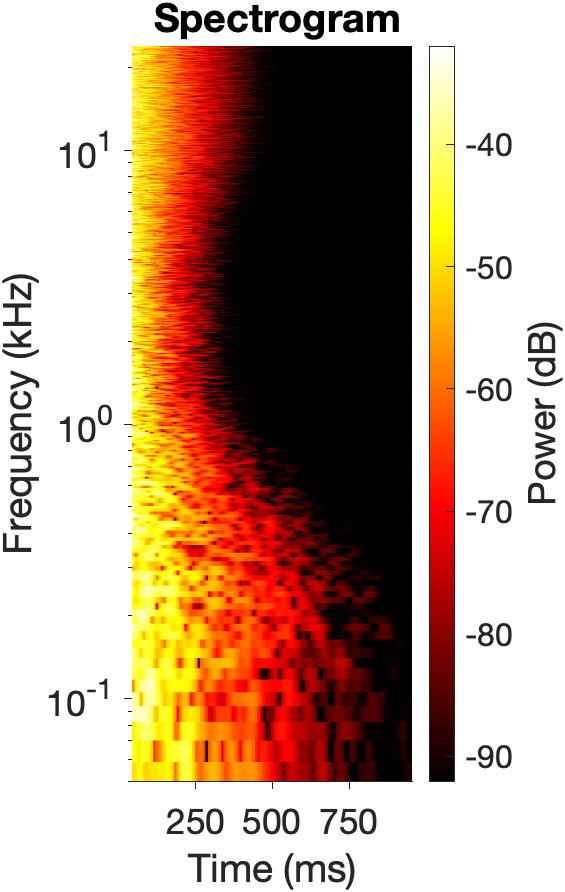}%
    }
      \caption{\label{FIG:SH_EXP_CONV}Sampled $T_{60}$ in Fig. \ref{FIG:SH_EXP_CONV:T60} revert to the prior mean function $\overline{T}_{60}(\omega, \theta, \phi) = 0.5$ when further from the observed $T_{60}(\omega, 90^\circ, 0^\circ)$. Exponentiating convolutions are shown for  varying $\phi$ on the horizontal plane $\theta = 90^\circ$.}    
\end{figure*}

\textbf{Directional Reverberation Generation:} 
%
Consider a normally distributed real-valued random pressure field $\MAT{P}$ represented by its first-order $(N_C=1)$ SH expansion $\VEC{C}$, and a GP distribution of direction dependent filters $\VEC{g}(\theta, \phi)$.
We specify a GP prior with the constant prior mean function $\overline{T}_{60}(\omega, \theta, \phi) = 0.5$ second in-place of Eq. \eqref{EQ:GPR_PRIOR_MU}, and non-stationary covariance of Eq. \eqref{EQ:GPR_COV} with hyper-parameters $\alpha = 0.5$, $\beta = 0$, $\sigma = 0.25$, $\gamma = 0.6$, $\ell = 343$.
Suppose we observe $\VEC{y} = T_{60}(\underline{\VEC{\omega}}, \underline{\VEC{\theta}}, \underline{\VEC{\phi}})$ at coordinates $\CBRAK{\underline{\VEC{\omega}}, \PAREN{\underline{\VEC{\theta}}, \underline{\VEC{\phi}}}} = \CBRAK{\CBRAK{0, 693, 24000} \textrm{Hz} \times (90^\circ, 0^\circ)}$ following Eq. \eqref{EQ:GPR_SUPPORT} as shown in Fig. \ref{FIG:SH_EXP_CONV:T60}. 
Conditioning the GP on the observations, we jointly sample $T_{60}(\VEC{\omega}, \VEC{\theta}, \VEC{\phi})$ at the predictive mean function $\VEC{\mu}_* (\VEC{X}_* = \CBRAK{\VEC{\omega} \times (\VEC{\theta}, \VEC{\phi}) } )$ in Eq. \eqref{EQ:GPR} over uniform frequencies $\VEC{\omega} \in \field{R}^{N}$ for $N = 16$, and uniform Fibonacci lattice \cite{hannay2004fibonacci} spherical coordinates $\VEC{\theta}, \VEC{\phi} \in \field{R}^{N_E}$, where $N_E = 25$ is the number of bases of a $4^{th}$-order SH expansion. 
The sampled $T_{60}(\VEC{\omega}, \VEC{\theta}, \VEC{\phi})$ encode our directional reverberation-time function as it veers from the observed $T_{60}(\underline{\omega}, \underline{\theta}, \underline{\phi})$ towards the prior mean $\overline{T}_{60}(\omega, \theta, \phi)$ for larger $\omega$ and $d(\theta, \phi, \underline{\theta}, \underline{\phi})$. 
Thus, we realize the mean reverberation field of the GP by optimizing direction dependent exponentiating filters $\VEC{g}(\VEC{\theta}, \VEC{\phi}) \in \field{R}^{N \times N_E}$ for design targets $T_{60}(\VEC{\omega}, \VEC{\theta}, \VEC{\phi})$ on the spherical coordinate grid via Eq. \eqref{EQ:NEAR_MIN_PHASE} for $\tau = 1$.

Applying our directional filters $\VEC{g}(\VEC{\theta}, \VEC{\phi})$ to pressure field $\VEC{P}$ in Eq. \eqref{EQ:DIR_REVERB_P}, we compute the exponentiating convolution pressure field matrix $\tilde{\MAT{P}}$ via Algorithm \ref{ALG:EXP_CONV:OPT} over the spherical grid $(\VEC{\theta}, \VEC{\phi})$, and re-expand along SH bases to yield the directional reverberation expansion matrix $\tilde{\MAT{E}} = \MAT{Y}^{\minus 1}(\VEC{\theta}, \VEC{\phi})  \tilde{\MAT{P}}$. For cross-validation, we evaluate the expansion $\tilde{\MAT{E}}$ at spherical coordinates on the horizontal plane outside the spherical grid $(\VEC{\theta}, \VEC{\phi})$ as shown in Figs. \ref{FIG:SH_EXP_CONV:SAMP_1}$-$\ref{FIG:SH_EXP_CONV:SAMP_4}. The resulting pressure/time spectrograms acquire the desired $T_{60}$ profiles of our predictive mean function at azimuth angles that span the nearest to furthest spherical coordinates from the observation. Reducing the grid size to $N_E = 16$ introduced spatial aliasing in high-frequency for small $T_{60}$ near $\phi = 0$.

\section{Conclusions}
\label{SEC:CONC}

We presented several new methods for both augmenting and generating SH-RIRs with time-varying and direction dependent reverberation. First, we generalized the convolution operator for time-varying exponentiating filters and derived two fast algorithms for their efficient computation. We then extended our methods to SH transforms of multi-mic pressure fields by showing that exponentiating convolution commutes with all linear transforms. Next, we introduced direction dependent filtering and reverberation over finite-order SH bases using smooth filters in frequency and spherical coordinates. Our second-order cone program realized near minimum-phase filters from smooth $T_{60}(\omega, \theta, \phi)$ spectral-decay functions sampled from non-stationary GPs. Experiments validated our process for generating directional reverberation fields from Gaussian noise.

\section{Appendix}

Let $\MAT{A} \in \field{C}^{N_A \times N_H}$ be a linear transform matrix mapping $N_H$ number of input to $N_A$ output channels. Multiplying $\MAT{A}$ by the results of the exponentiating convolution in Eq. \eqref{EQ:EXP_CONV} yields the matrix $\tilde{\MAT{F}} = \MAT{A} \MAT{F}  \in \field{C}^{N_A \times MN}$ where $\MAT{F} = \BRAK{\VEC{f}_1, \hdots, \VEC{f}_{N_A} }^T \in \field{R}^{N_H \times MN} $ is the row-vector matrix of exponentiating convolutions $\VEC{f}_k = f(\VEC{h}_k, \VEC{g}) \in \field{R}^{MN \times 1} $. The entrants of $\tilde{\MAT{F}}$ are therefore given by
\begin{equation} \label{EQ:COMMUTE_EXP_POST}
\displaystyle
\begin{split}
\tilde{F}_{in} & = \sum_{k=1}^{N_H}  A_{ik} \VEC{f}_k[n] \qquad \textrm{Post-convolution} \\ 
 & = \sum_{k=1}^{N_H} \sum_{m = 1}^{MN} A_{ik}  h_k[n - m + 1] \, g(n -  m + 1) [m].
\end{split}
\end{equation}
Conversely, multiplying $\MAT{A}$ by the multichannel matrix $\MAT{H}$ in Eq. \eqref{EQ:MULTICH} before the exponentiating convolution yields $\tilde{\MAT{H}}  = \MAT{A} \MAT{H} \in \field{C}^{N_A \times M}$ where  $\tilde{\VEC{h}}_i$ is the $i^{th}$ row of $\tilde{\MAT{H}}$, and $\tilde{H}_{in} = \sum_{k=1}^{N_H} A_{ik} h_k[n]$. Evaluating the exponentiating convolution over  $\tilde{\VEC{h}}_i$ in  $f(\tilde{\VEC{h}}_i, \VEC{g})$  gives
\begin{equation} \label{EQ:COMMUTE_EXP_PRE}
\displaystyle
\begin{split}
f_i[n] & = \sum_{m = 1}^{MN} \tilde{h}_i[n - m + 1] \, g(n -  m + 1) [m] \quad \textrm{Pre-conv.} \\
& = \sum_{m = 1}^{MN} \sum_{k=1}^{N_H} A_{ik} h_k[n - m + 1]  \, g(n -  m + 1) [m],
\end{split}
\raisetag{8ex}
\end{equation}
which is equivalent to Eq. \eqref{EQ:COMMUTE_EXP_POST}. Analogous substitutions of linear transformation orderings into recursive convolutions of Eq. \eqref{EQ:REC_CONV} give the same equivalences.

In the case of linear transforms of the SH product operator $A(\MAT{D})$ expressed by the transfer matrix $\MAT{A}$, its $n^{th}$ column has entrants specified along the SH bases with degree $L$ bounded between $0 \leq L \leq L_D + L_C$, and order $M$ bounded between $-L \leq M \leq L$  as follows:
\begin{equation} \label{EQ:SH_PROD_MAT_A}
\displaystyle
\begin{split}
A_L^M (n) & = \sum_{l = 0}^{L_D }  \, \sum_{k = \underline{k}}^{ \overline{k} }  \sqrt{\frac{(2 l + 1) (2 k + 1) }{4 \pi (2L + 1)} }  \,  c \PAREN{\begin{array}{ccc} l & k & L \\ 0 & 0 & 0 \end{array}}  \\
&\times   \sum_{m =\underline{m} }^{\overline{m}}   c \PAREN{\begin{array}{ccc} l & k & L \\ m & M - m &  M \end{array}} D_{l}^{m} \, \mathds{1}(n) ,   \\
\mathds{1}(n) & = \left \{ \begin{array}{cc}1, &  n = l^2 + M - m + k + 1 \\ 0, &  \textrm{Otherwise} \end{array}\right . , \\
\end{split}
\raisetag{6ex}
\end{equation}
where $c(*)$ are the Wigner's 3-j symbols \cite{messiah1962clebsch}, and $\mathds{1}(n)$ is an indicator function of column index $n$. The summations indexes $k$ are bounded between $\underline{k}  = \ABS{L - l}$,  $\overline{k}  = \min(\ABS{L + l},  \,  L_C )$, and  $m$ between $\underline{m} =  \max(M - k,   \, -l)$, $\overline{m} = \min(M + k, \,  l)$.



\bibliographystyle{jaes}

\bibliography{refs}

@inproceedings{hamilton2021air,
  title={Air absorption filtering method based on approximate green's function for Stokes' equation},
  author={Hamilton, Brian},
  booktitle={2021 24th International Conference on Digital Audio Effects (DAFx)},
  pages={160--167},
  year={2021},
  organization={IEEE}
}

@article{kates2020adding,
  title={Adding air absorption to simulated room acoustic models},
  author={Kates, James M and Brandewie, Eugene J},
  journal={The Journal of the Acoustical Society of America},
  volume={148},
  number={5},
   IGNOREpages={EL408--EL413},
  year={2020},
  publisher={AIP Publishing}
}

@inproceedings{pfaff2017filtering,
  title={Filtering on the unit sphere using spherical harmonics},
  author={Pfaff, Florian and Kurz, Gerhard and Hanebeck, Uwe D},
  booktitle={2017 IEEE International Conference on Multisensor Fusion and Integration for Intelligent Systems (MFI)},
  pages={124--130},
  year={2017},
  organization={IEEE}
}

@article{luo2021spherical,
  title={Spherical harmonic covariance and magnitude function encodings for beamformer design},
  author={Luo, Yuancheng},
  journal={EURASIP Journal on Audio, Speech, and Music Processing},
  volume={2021},
  number={1},
  pages={41},
  year={2021},
  publisher={Springer}
}

@PhDThesis{jarosz08thesis,
    title = "Efficient Monte Carlo Methods for Light Transport in Scattering Media",
    author = "Wojciech Jarosz",
    year = "2008",
    month = "September",
    school = "UC San Diego"
}

@inproceedings{hold2021spatial,
  title={Spatial filter bank in the spherical harmonic domain: Reconstruction and application},
  author={Hold, Christoph and Schlecht, Sebastian J and Politis, Archontis and Pulkki, Ville},
  booktitle={2021 IEEE Workshop on Applications of Signal Processing to Audio and Acoustics},
  year={2021},
}

@article{messiah1962clebsch,
  title={Clebsch-Gordan (C.-G.) coefficients and'3j'symbols},
  author={Messiah, A},
  journal={Appendix CI in Quantum Mechanics},
  volume={2},
  pages={1054--1060},
  year={1962},
  publisher={North-Holland Amsterdam, The Netherlands}
}

@article{zotter2012all,
  title={All-round ambisonic panning and decoding},
  author={Zotter, Franz and Frank, Matthias},
  journal={Journal of the audio engineering society},
  volume={60},
  number={10},
  pages={807--820},
  year={2012},
  publisher={Audio Engineering Society}
}

@article{suda2002fast,
  title={A fast spherical harmonics transform algorithm},
  author={Suda, Reiji and Takami, Masayasu},
  journal={Mathematics of computation},
  volume={71},
  number={238},
  pages={703--715},
  year={2002}
}

@article{paciorek2003nonstationary,
  title={Nonstationary covariance functions for Gaussian process regression},
  author={Paciorek, Christopher and Schervish, Mark},
  journal={Advances in neural information processing systems},
  volume={16},
  year={2003}
}

@incollection{rasmussen2003gaussian,
  title={Gaussian processes in machine learning},
  author={Rasmussen, Carl Edward},
  booktitle={Summer school on machine learning},
  pages={63--71},
  year={2003},
  publisher={Springer}
}

@article{pei2006minimum,
  title={Minimum-phase FIR filter design using real cepstrum},
  author={Pei, S-C and Lin, H-S},
  journal={IEEE Transactions on Circuits and Systems II: Express Briefs},
  volume={53},
  number={10},
  pages={1113--1117},
  year={2006},
  publisher={IEEE}
}

@book{oppenheim1999discrete,
  author    = {Oppenheim, Alan V. and Schafer, Ronald W. and Buck, John R.},
  title     = {Discrete-Time Signal Processing},
  edition   = {2nd},
  publisher = {Prentice Hall},
  year      = {1999},
  address   = {Upper Saddle River, NJ},
}

@article{hannay2004fibonacci,
  title={Fibonacci numerical integration on a sphere},
  author={Hannay, John Howard and Nye, John Frederick},
  journal={Journal of Physics A: Mathematical and General},
  volume={37},
  number={48},
  pages={11591--11601},
  year={2004}
}

@book{rafaely2015fundamentals,
  title={Fundamentals of spherical array processing},
  author={Rafaely, Boaz},
  volume={8},
  year={2015},
  publisher={Springer}
}

@book{muller2006spherical,
  title={Spherical harmonics},
  author={M{\"u}ller, Claus},
  year={2006},
  publisher={Springer}
}

@book{jarrett2017theory,
  title={Theory and applications of spherical microphone array processing},
  author={Jarrett, Daniel P and Habets, Emanu{\"e}l AP and Naylor, Patrick A},
  volume={9},
  year={2017},
  publisher={Springer}
}

@book{zotter2019ambisonics,
  title={Ambisonics: A practical 3D audio theory for recording, studio production, sound reinforcement, and virtual reality},
  author={Zotter, Franz and Frank, Matthias},
  year={2019},
  publisher={Springer}
}

@inproceedings{wang2023time,
  title={Time-domain wideband image source method for spherical microphone arrays},
  author={Wang, Jiarui and Zhang, Jihui Aimee and Samarasinghe, Prasanga and Abhayapala, Thushara},
  booktitle={2023 IEEE 25th International Workshop on Multimedia Signal Processing (MMSP)},
  pages={1--6},
  year={2023},
  organization={IEEE}
}

@article{xu2024simulating,
  title={Simulating room transfer functions between transducers mounted on audio devices using a modified image source method},
  author={Xu, Zeyu and Herzog, Adrian and Lodermeyer, Alexander and Habets, Emanu{\"e}l AP and Prinn, Albert G},
  journal={The Journal of the Acoustical Society of America},
  volume={155},
  number={1},
  pages={343--357},
  year={2024},
  publisher={AIP Publishing}
}

@INPROCEEDINGS{luo2021FSRR,
  author={Luo, Yuancheng and Kim, Wontak},
  booktitle={2020 28th European Signal Processing Conference (EUSIPCO)}, 
  title={Fast Source-Room-Receiver Acoustics Modeling}, 
  year={2021},
  volume={},
  number={},
  pages={51-55}}

@article{valimaki2017late,
  title={Late reverberation synthesis using filtered velvet noise},
  author={V{\"a}lim{\"a}ki, Vesa and Holm-Rasmussen, Bo and Alary, Benoit and Lehtonen, Heidi-Maria},
  journal={Applied Sciences},
  volume={7},
  number={5},
  pages={483},
  year={2017},
  publisher={MDPI}
}

@inproceedings{fagerstrom2022dark,
  title={Dark velvet noise},
  author={Fagerstr{\"o}m, Jon and Meyer-Kahlen, Nils and Schlecht, Sebastian J and V{\"a}lim{\"a}ki, Vesa},
  booktitle={International Conference on Digital Audio Effects},
  pages={192--199},
  year={2022},
  organization={DAFx}
}

\end{document}